# Dynamic Hysteresis Probes High-β Nanolaser Emission Regimes

Si Hui Pan[1*], Qing Gu[1†], Abdelkrim El Amili[1], Felipe Vallini[1], Yeshaiahu Fainman[1]

**Affiliations:**

[1]Department of Electrical and Computer Engineering, University of California at San Diego, La Jolla, CA, 92093-0407, USA.

*Correspondence to: fainman@ece.ucsd.edu.

†Current addresses: Department of Electrical Engineering, University of Texas at Dallas, Richardson, TX, 75080-3021, USA.

**Abstract**:

The quest for an integrated light source that promises high energy efficiency and fast modulation for high-performance photonic circuits has led to the development of room-temperature telecom-wavelength nanoscale laser with high spontaneous emission factors, β. The coherence characterization of this type of lasers is inherently difficult with the conventional measurement of output light intensity versus input pump intensity due to the diminishing kink in the measurement curve. We demonstrate the transition from chaotic to coherent emission of a high-β pulse-pump metallo-dielectric nanolaser can be determined by examining the width of a second order intensity correlation, $g^2(\tau)$, peak, which shrinks below and broadens above threshold. Photon fluctuation study, first one ever reported for this type of nanolaser, confirms the validity of this measurement technique. Additionally, we show that the width variation above threshold results from the delayed threshold phenomenon, providing the first indirect observation of dynamical hysteresis in a nanolaser.

**Main Text:**

Nanocavity lasers with high spontaneous emission factors, β, have attracted considerable attention in the past decade in light of their technical applications ranging from optical interconnects (*1*), bio-sensing (*2*), chemical detection (3) and nonlinear optical microscopy (*4*) to fundamental research on thresholdless lasers (*5-8*) and cavity quantum electrodynamics (*9-10*). Theoretically a high-β nanolaser is more energy efficient as most spontaneous emission is funneled into the lasing mode, resulting in an extremely low lasing threshold (*5*). Additionally, the spontaneous emission rate can be significantly enhanced due to Purcell effect in high-β nano-resonators, pushing the fundamental modulation speed limits of on-chip coherent sources to beyond 100 GHz (*10*). A wide spectrum of resonator architectures has been explored, including photonic crystal cavities (*7-8,11-14*), circular (*15*) and rectangular (*16*) metallo-dielectric cavities, metal-semiconductor coaxial cavities (*6*), nanopillar/nanowire cavities (*17-18*) and plasmonic cavities (SPASERs) (*19-22*). To date, the majority of studies on nanocavity lasers have focused on the proof of concept demonstration of lasing behavior. In particular, the measurement of a "kink" in the light-out versus light-in curve (LL-curve) is prevalently used as affirming evidence for lasing and is often the only method adopted to identify the lasing threshold (*6-7, 12-13, 15-21*). However, it is well known that the "kink" in the LL-curve diminishes as β increases. Therefore, for high-β nanocavity lasers, LL-curve measurement alone can neither fully characterize lasing nor unambiguously identify the lasing threshold.

On the other hand, measuring the second order intensity correlation function $g^2(\tau) = \left\langle I(t) \right\rangle \left\langle I(t+\tau) \right\rangle \big/ \left\langle I(t) \right\rangle^2$, where $\left\langle I(t) \right\rangle$ represents the expectation value of the nanolaser output intensity at time $t$, is a more definitive method to confirm lasing (*8, 11, 14, 22*). The observation of a photon-bunching peak around zero delay ($\tau = 0$) signifies chaotic light emission and hence, its suppression univocally marks the onset of coherent emission. In contrast to the

LL-curve measurement, only a few $g^2(\tau)$ studies have been conducted with nanolasers due to their minuscule output power and relatively short coherence time. These constraints demand ultrasensitive photodetectors operating in the single photon counting regime with excellent timing precision. Commercial silicon-based single photon detectors can meet these demands. Hence, most $g^2(\tau)$ studies so far have focused on nanoscale lasers emitting at wavelengths below 1 µm. Furthermore, these studies are conducted at cryogenic temperatures under continuous wave pumping, with either the family of photonic crystal cavity lasers (*8, 11*) or plasmonic nanorod lasers (*22*). With telecommunication wavelengths, however, $g^2(\tau)$ measurements become more challenging due to poorer detector time resolution, which is detrimental to the photon-bunching peak. Optical pulse pumping typically employed for nanolasers further complicates the measurement because the photon-bunching peak overlaps with only the zero-delay pulse. Therefore, difference in height between the zero-delay and non-zero-delay pulses is the only indication that the bunching peak exists while its specific shape usually cannot be observed. Further, optical misalignment, signal-to-noise ratio and other measurement artifacts, such as detector crosstalk (*23*), can also affect the height difference, compromising its accuracy.

In this study, we show that by employing nanosecond pump pulses, different emission regimes of a nanolaser can be characterized via the full width at half maximum (FWHM) of a $g^2(\tau)$ peak. We demonstrate the applicability of this technique with a high-β metallo-dielectric nanolaser and further report the first coherence measurement on this type of nanolaser operating at room temperature. We demonstrate that while $g^2(\tau)$ peaks narrow in the spontaneous emission (SE) and amplified spontaneous emission (ASE) regimes as stimulated emission increases in proportion, their FWHMs increase in the lasing regime due to dynamical hysteresis

(DH), a phenomenon triggered when a laser is driven across its threshold. The usage of nanosecond pulses is the key to the observation of DH and, therefore, essential to this characterization technique. Previous $g^2(\tau)$ studies utilizing picosecond (*14*) or femtosecond (*24*) pulses could not observe this phenomenon because these pulse widths are far below the photodetector time resolution typically on the order of hundreds picoseconds.

The metallo-dielectric nanocavity design and fabrication techniques used in this study follows that of Nezhad *et al* (*15, 25*) (Fig. 1A). Scanning electron microscope (SEM) images of the device after dry etching and silver sputtering are shown in Fig. 1B and 1C, respectively. The measured spectral evolution of the device from below to above threshold is presented in Fig. 1D. Broadband photoluminescence (PL) spectra at low pump power indicate the emission is predominantly spontaneous. As the pump power increases, a narrow spectral line appears at around 1421nm, corresponding to the lasing mode. The integrated output power from the PL spectra as a function of input pump power in a log-log scale (LL-curve) is presented in Fig. 2A. Using a rate-equation model (*25*) with a time-dependent pump pulse and β as a free parameter, we were able to fit the experimental LL-curve and obtain a β factor of 0.25. The different operational regimes of our nanolaser, including SE, ASE and stimulated emission, are clearly identified in the LL-curve. The threshold power, $P_{th}$, as conventionally defined by the kink in the LL-curve, is approximately 200 μW.

To further verify the nano-device is indeed a laser, we first examined the normalized intensity correlation at zero delay, $g^2(0)$, as a function of pump power (Fig. 2B). The points labeled as I, II and III in Fig. 2B were calculated from the experimental non-normalized correlation histograms shown in Fig. 3A(D), 3B(E) and 3C(F), respectively. Regardless of the output photon statistics of our device, the non-normalized correlation histograms are expected to have pulses at a frequency equivalent to the pump laser repetition rate, as depicted in Fig. 3A, B, C. In addition, the output emission of the device exhibits chaotic light characteristics [ $g^2(\tau = 0) > 1$ ] before lasing occurs. Hence, in the SE and ASE regimes, extra correlation counts due to photon bunching will be added to the zero-delay pulse, making it taller than the non-zero-delay pulses (Fig. 3B, E). As the nanolaser transitions from below to above threshold, the emitted photons statistics evolves from super-Poissonian to Poissonian. Therefore, the extra bunching peak disappears and the zero-delay pulse height approaches that of a non-zero-delay pulse (Fig. 3C, F), indicating coherent light emission [ $g^2(\tau = 0) = 1$ ]. Careful examination of Fig. 2B shows that $g^2(0)$ reaches unity within measurement error when the pump power is equal to or greater than 600 μW or three times $P_{th}$. Therefore, in contrast to the popular intuition that lasing occurs once the kink in the LL-curve is observed, $g^2(\tau)$ measurement reveals that fully coherent light emission is achieved at pump power far beyond not only the kink but also the ASE region define by the nonlinear increase in the LL-curve. Similar results have been reported for lasers based on micropillar cavities (*26*) and photonic crystal cavities (*8, 14*). The slight difference between $g^2(0)$ and unity at $P > 600$ μW, as observed in Fig. 2B, is most likely due to a small amount of incoherent light emitted into a non-lasing mode at 1414nm (see Fig. 1D). Meanwhile, Fig. 3A, D shows that the photon-bunching peak also disappears far below threshold instead of approaching the theoretical value of two [ $g^2(\tau = 0) = 2$ ] expected for an ideal thermal light source. This is because the coherence time, $\tau_c$, is extremely short compared to the time resolution of the photodetectors (~100 ps) when SE dominates. The width of the photon bunching peak is on the order of the coherence time (*27*). Thus, as $\tau_c$ decreases, averaging effect due to detector timing uncertainty gradually washes out the "fingerprint" of the super-Poissonian statistics below threshold.

While the pulse height difference has been previously observed and used to demonstrate lasing (*14, 24*), we show here the very first detail investigation of the pulse width variation at different emission regimes of a nanolaser. Comparisons between Fig. 3D, E and F indicate that the $g^2(\tau)$ pulses are widest in the SE regime while the width variations in the ASE and lasing regimes are more subtle. We extract the $g^2(\tau)$ pulse FWHM from the experimental correlation histogram and plot it as a function of pump power from far below to far above threshold in Fig. 2C. As shown, the pulses narrow with increasing pump power in the SE regime and reach a minimum in the ASE regime. The pulses then broaden with further increasing pump power in the lasing regime until more than four times of $P_{th}$ when pulse narrowing is observed again. The zero-delay pulse is narrower than the non-zero-delay pulses around threshold due to an extra photon bunching peak as discussed above. Given the large delay-time intervals between adjacent pulses ($\Delta\tau \sim 3\mu s$) in $g^2(\tau)$, change in photon statistics described above affects only the zero-delay pulse and therefore, cannot account for the pulse narrowing or broadening behaviors which apply to both non-zero-delay and zero-delay pulses. Meanwhile, the distinct evolutional trends in the FWHM before and after lasing occurs indicate their close relationship to the nanolaser operational regimes. To ensure the observed FWHM variation is not caused by modifications in the pump laser itself, we also characterize the FWHM of the pump laser $g^2(\tau)$ as a function of pump power  (Fig. 2C). The results, which are taken over several hours and repeatable on a daily basis, reveal negligible variation in the pump laser $g^2(\tau)$ width, confirming the pulse narrowing and broadening effects result primarily from the nanolaser operation mechanism.

To further understand the mechanism governing the pulse width evolution, we employ a rate-equation model to examine the output photon and carrier densities as functions of time (*25*). Fig. 4 shows the theoretical responses of the nanolaser subjected to a time-dependent Gaussian pump pulse at peak power far below threshold (Fig. 4A), slightly below threshold (Fig. 4B), slightly above threshold (Fig. 4C) and far above threshold (Fig. 4D, E). By taking the autocorrelation (*25*) of the photon pulses shown in Fig. 4A-E, the $g^2(\tau)$ functions are generated numerically and shown in Fig. 4F-J. The FWHM of the simulated $g^2(\tau)$ pulse as a function of pump power is plotted in the inset of Fig. 2C. In the SE regime, the width of the output optical pulse is primarily determined by the pump pulse width (Fig. 4A), the photon lifetime ($\tau_p$) and the radiative recombination lifetime ($\tau_{rc}$), which can be approximated by the SE lifetime in this far below threshold region. As the peak pump power increases, stimulated emission grows in contribution and the device gradually transitions to the ASE regime. Stimulated emission accelerates radiative recombination in the cavity, which reduces the width of the output optical pulse (Fig. 4B). This pulse-shortening phenomenon contributes to the narrowing effect in the SE and ASE regions of Fig. 2C. Additionally, the narrowing rate here is significantly influenced by $\tau_{rc}$, which is reduced due to the prominent Purcell effect in nanocavity lasers (*9-10*). We found a satisfactory agreement between theory and experiment when $\tau_{rc} = 2$ ns (Fig. 2C and inset). Comparison between this value and the radiative recombination lifetime of a MQWs wafer, where $\tau_{rc} \approx 100$ ns (*28*), gives a Purcell enhancement factor of approximately 50, which is consistent to the same order of magnitude with the value, 53, estimated by the method used in ref. 9 and 15.

As the peak pump power continues to increase, the FWHM reaches a minimum at around threshold and begins to broaden. Such broadening effect is a consequence of delayed threshold phenomenon (DTP), which arises when any nonlinear system crosses a bifurcation point between two stable solution branches (*29-30*). In a laser, the bifurcation point corresponds to the lasing threshold and DTP manifests itself in the form of a jump in the output intensity when the pump intensity is swept across threshold. Such phenomenon has been previously observed in macroscopic lasers (*31-33*). Fig. 4C, D and E show that DTP is also present in our nanolaser. As the time dependent Gaussian pump pulse sweeps across the nanolaser threshold $P_{th}$ at time $t_1$ (Fig. 4C, D, E), the nonlinear system does not respond immediately due to the infinite response time at the bifurcation point, $P_{th}$ (*30*). Physically stimulated emission cannot happen instantaneously. The time it takes to generate enough stimulated photons to statistically dominate the emission depends on the nanolaser characteristic times, such as photon and carrier lifetimes. Hence, the nanolaser remains in its nonlasing state for some time after the pump pulse intensity crosses $P_{th}$. When stimulated emission finally builds up in the cavity such that the nanolaser switches to its lasing state, the output photon intensity increases by more than an order of magnitude in a very short time period (~1 ns). This fast change in intensity truncates part of the output pulse on the left-hand side, leading to an asymmetrical pulse shape. The missing parts of the photon pulses are represented by blue shaded regions in Fig. 4C, D, E. The nanolaser remains in the lasing state for as long as $P(t) > P_{th}$. However, once the pump power reduces to below $P_{th}$ at time $t_2$, lasing ceases and the device gradually decays back to its initial state. Unlike the transition from nonlasing to lasing, the reverse process happens smoothly because the cavity is already filled with photons whose number gradually decays to zero after $t_2$. The decaying process happens continuously. Therefore, no sudden change in intensity is observed on the right-hand side of the output pulse. This asymmetry between the response of the laser when the pump is swept from below to above threshold and vice versa is often termed as dynamical hysteresis in the literature (*29-30, 33*).

Above threshold, a higher peak pump power level corresponds to a smaller $t_1$ (Fig. 4C, D, E). Therefore, the intensity jump on the left-hand side of the output pulse happens at an earlier time, which trims away less of the photon pulse, as indicated by the reducing areas of the blue shaded regions with increasing peak pump power in Fig. 4C, D, E. This is consistent with the fact that the time delay associated with DTP reduces as the first derivative of the pump pulse increases proportionally to the peak pump power (see Fig. S1-S3) (*25, 29-33*). Additionally, as the pump pulse stays above threshold for a longer period of time, the device stays lasing for longer and the photon decay on the right-hand side of the output pulse happens at a later time. That is $\Delta t = t_2 - t_1$ increases since $t_1$ decreases while $t_2$ increases. All these factors contribute to a longer output pulse, which leads to broadening of the $g^2(\tau)$ functions as the peak pump power increases above threshold (Fig. 2C). Theoretically, at very high peak power, $\Delta t = t_2 - t_1$ approaches infinity and nearly the entire pump pulse is above threshold. At this point, the output pulse mirrors the pump and its width is determined entirely by the width of the pump pulse. However, in order to reach this condition, the peak power must be at least a hundred times above threshold (see Fig. S4) (*25*), which is unachievable in practical experiments.

In fact, as shown in Fig. 2C, the experimental curve deviates far from theory once the peak pump power becomes more than 600 µW or three times the threshold power. Such discrepancy can be explained by thermal saturation of the device at high pump power, which is unaccounted for in our rate-equation model. Above threshold, gain material properties degrade as a result of rapid internal temperature rise in the device with increasing pump power, which reduces the device internal quantum efficiency. In addition, metal losses are proportional to temperature. Therefore, the quality factor is expected to reduce as a result of self-heating. In combination these effects cause an increase in the nanolaser threshold (*34-35*), which is effectively equivalent to reducing the pump (or injected) carrier density. Therefore, self-heating of the device leads to a reduction in the $g^2(\tau)$ pulse width far above threshold. Consequently, the above-threshold behavior of the $g^2(\tau)$ FWHM depends on the interplay between broadening due to DH and narrowing due to self-heating. Below three times the threshold, $P < 3P_{th} = 600$ µW, width broadening due to DH dominates and the experimental data follow the same trend as predicted by theory (Fig. 2C inset). Above three times the threshold, corresponding to when the nanolaser emission is fully coherent, narrowing effect due to self-heating balances the DTP induced broadening and the experimental FWHM stays approximately constant. Beyond five times the threshold power, self-heating dominates and pulse width narrowing is again observed for the last three pump power levels. Thermal saturation also causes roll-over in the LL-curve at high pump power (*34*), which predicts the measured laser output intensities to be less than the theoretical values given by our rate-equation model. Such difference has been observed in our experimental data (Fig. 2A), confirming the existence of self-heating at high pump power.

In summary, we examined the $g^2(\tau)$ pulse width as a function of pump power and demonstrated the width evolution is closely linked to the different operational regimes of our high-β metallo-dielectric nanolaser whose lasing behavior has been confirmed by both LL-curve and second order intensity correlation at zero delay. We demonstrated that the $g^2(\tau)$ pulses shrink in width below and near threshold and broaden above threshold when nanosecond pump pulses are employed. It is, therefore, feasible to exploit such a measurement technique to identify the SE, ASE and lasing regimes of a high-β nanolaser when its LL-curve fails to be reliable due to its diminishing kink. By exploring a rate-equation model, we discover that the modified SE lifetime of our nanocavity heavily influences the below threshold narrowing rate of the $g^2(\tau)$ pulses, offering valuable information about the Purcell factor which neither the LL-curve nor the $g^2(0)$ measurement could provide. On the other hand, dynamical hysteresis and self-heating effect determine the above threshold behavior of the $g^2(\tau)$ pulse width. While dynamical hysteresis in macroscopic lasers has been studied before, this is the first time such nonlinear dynamics has been observed in a nanolaser through the second order intensity correlation measurement. Our work demonstrates that second order intensity correlation measurement conducted under pulse pumping in the nanosecond time scale can be utilized to extract information about the nonlinear dynamics and Purcell factor of nanolasers besides confirming Poisson statistics of their output. This work paves the way for future investigations of the unexplored physics of nonlinear dynamics in ultra-small resonators.

**Acknowledgments:** This research was supported by the National Science Foundation (NSF), the NSF Center for Integrated Access Networks, the Office of Naval Research (ONR) Multi-Disciplinary Research Initiative, the ONR Army Research Office and the Cymer Corporation. S. H. Pan was supported by the National Science Foundation Graduate Research Fellowship under Grant No. DGE-1144086.

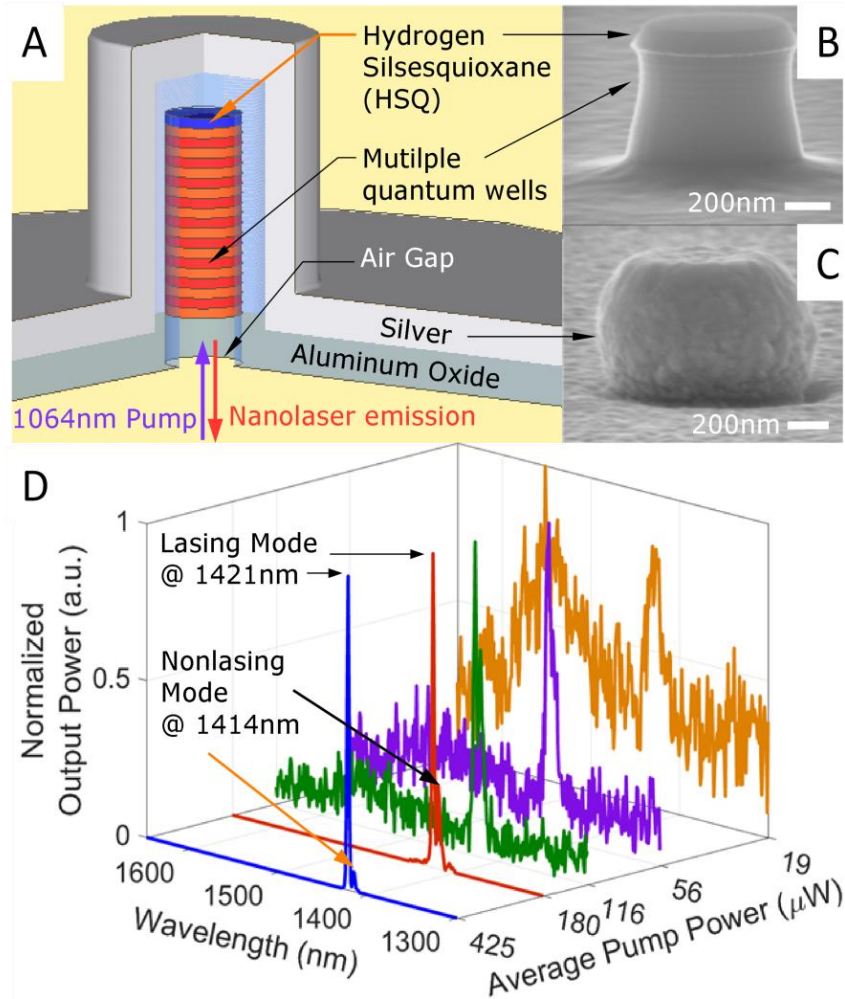

**Fig. 1.** (A) A schematic of the fabricated metallo-dielectric nanolaser: a 300nm of multiple quantum wells gain medium is surrounded by 100nm of oxide, which is covered by 300nm of silver. (B) - (C) SEM images of the nanolaser after reactive ion etching (B) and silver deposition (C). (D) Normalized spectral evolution of our high-β nanolaser indicates a broadband spontaneous emission spectrum at low pump power and a distinct lasing peak (1421nm) accompanied by a non-lasing mode (1414nm) at high pump power.

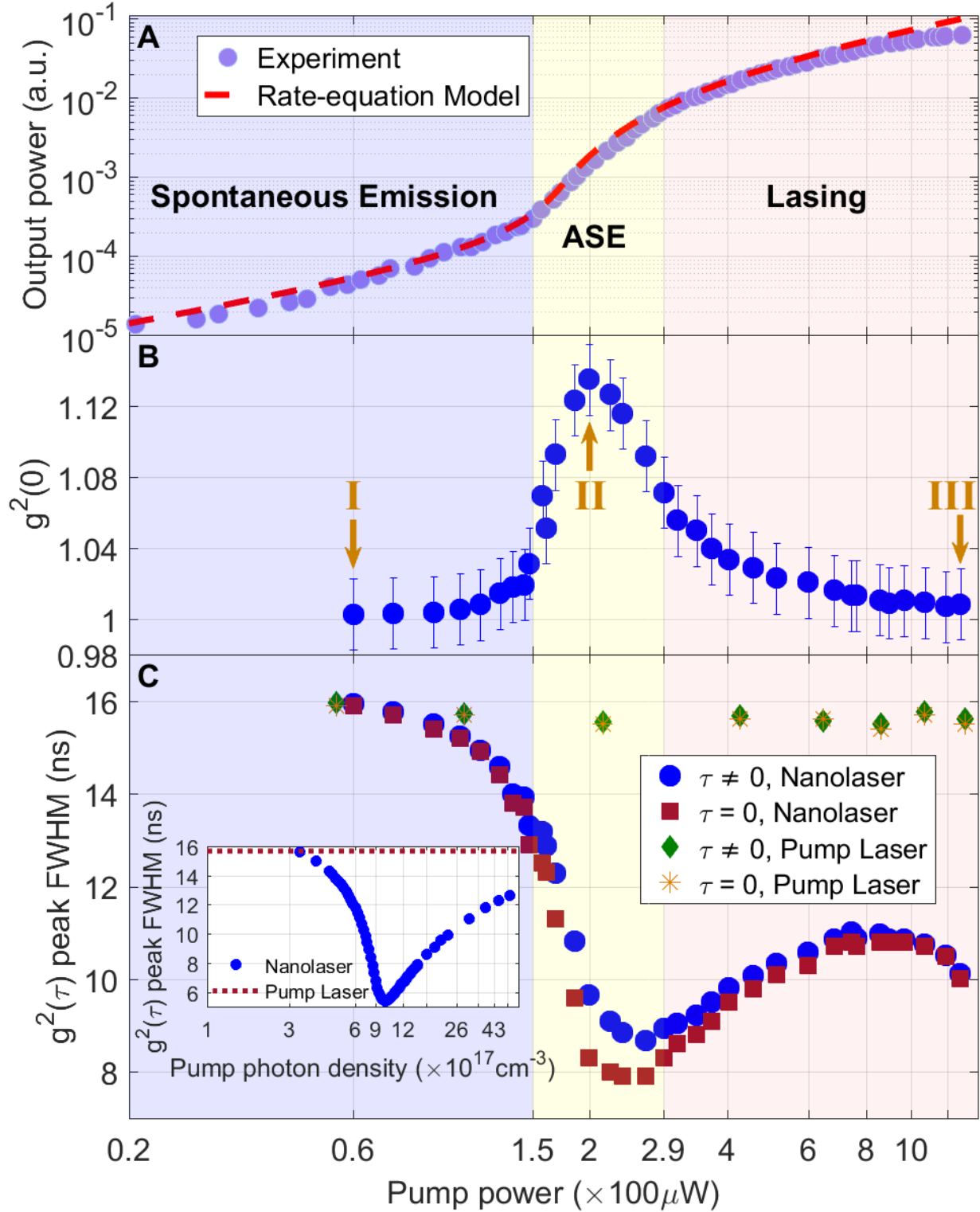

**Fig. 2.** (A) LL-curve: experimental (circles) and theoretical (dashed line) output power of the nanolaser as functions of input pump power. The theoretical LL-curve is generated by fitting the experimental data with a rate-equation model. The SE factor $\beta$ is fitted to be 0.25. (B) Evolution of the nanolaser's second order intensity correlation function at zero delay, $g^2(0)$, which confirms lasing as it approaches unity at high pump power. It decays at low pump power because the coherence time drops below the detection limit of our setup. (C) Experimental FWHM of the nanolaser and pump laser $g^2(\tau)$ pulses as functions of the pump power. FWHM narrows below threshold due to increasing radiative recombination rate as stimulated emission increases in fraction. Above threshold, DTP or DH leads to broadening of the FWHM until self-heating dominates beyond 1 mW of pump power, at which point width narrowing is again observed. The inset plots the theoretical FWHM simulated with a rate-equation model without considering self-heating.

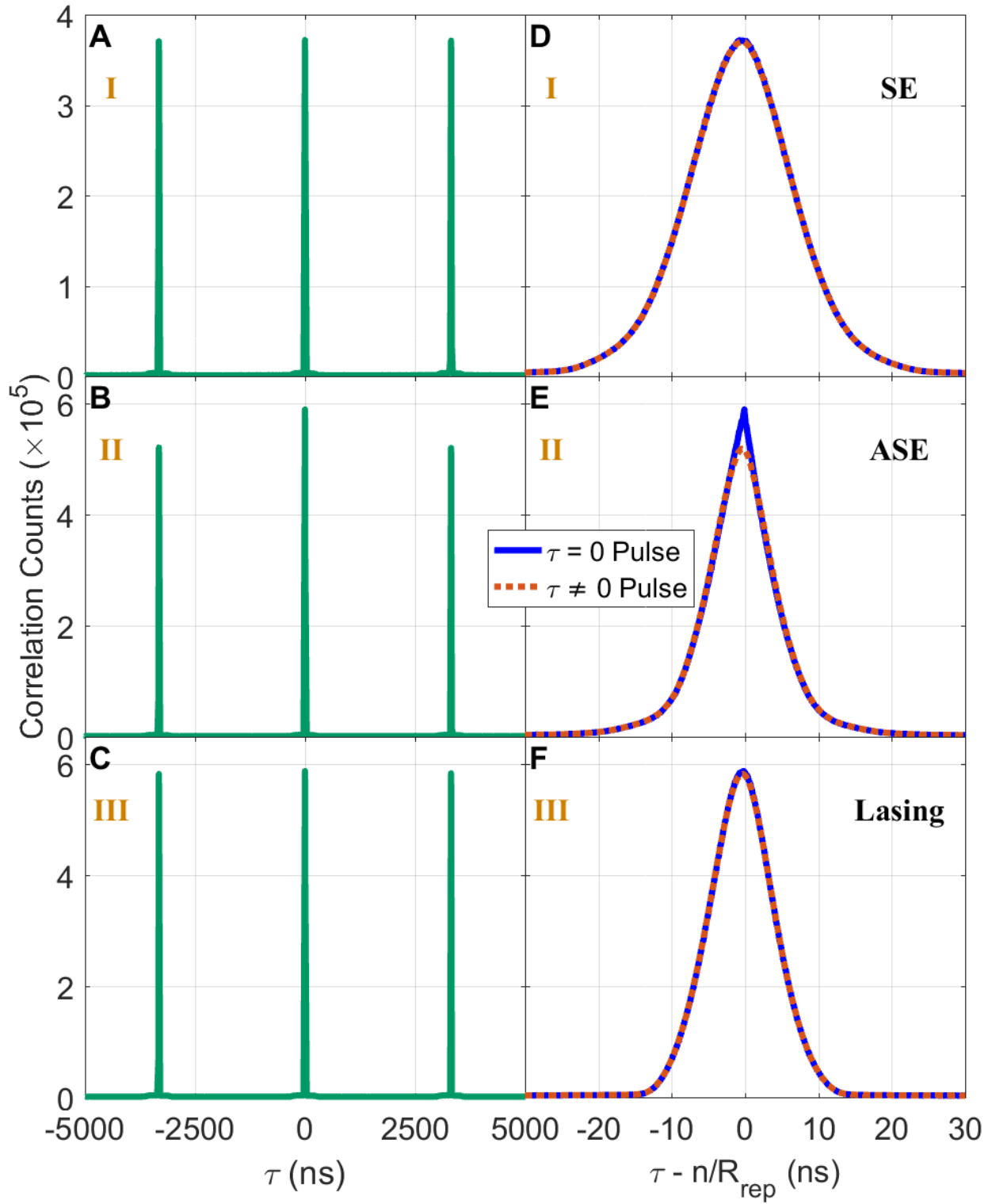

**Fig. 3.** (A) - (C) Experimental non-normalized intensity correlation histograms at pump power

levels labeled as I, II and III in Fig. 2B, corresponding to the SE, ASE and lasing regimes of the nanolaser. (D) - (F) are the same as (A) - (C) but with the non-zero-delay pulses overlaid on top of the zero-delay pulses. In the SE regime (A, D), the coherence time of the source is too short to be resolved by the photodetectors, resulting in the disappearance of the photon bunching peak. In the ASE regime (B, E), the coherence time lengthens and the photon bunching peak emerges, indicating the emission is partially chaotic. The photon bunching peak disappears (C, F) eventually when the nanolaser emission becomes fully coherent at high pump power. Additionally, the pulse FWHM varies at different emission regimes, with the broadest width appearing in the SE regime.

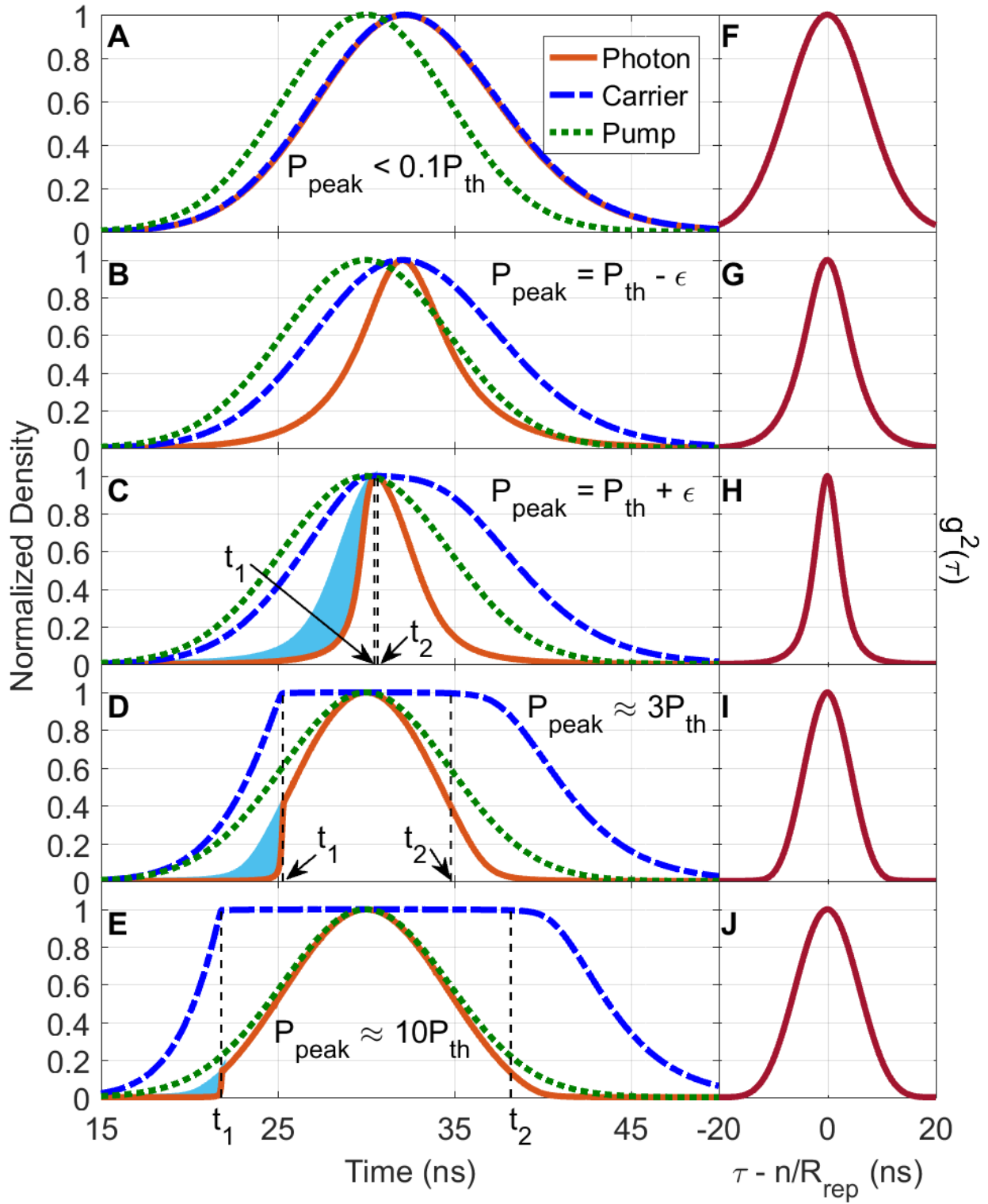

**Fig. 4.** (A) - (E) Normalized output photon, carrier and pump (injection) photon densities as functions of time at peak pump power, $P_{peak}$, far below threshold (A), slightly below threshold

(B), slightly above threshold (C) and far above threshold (D, E). Black dashed lines and labels $t_1$ and $t_2$ in (C), (D) and (E) indicate when the pump pulse crosses $P_{th}$. The output photon pulse follows the shape of the pump pulse in the SE regime (A), but narrows as the pump power increases (B). In the far below threshold regime, the narrowing rate is predominantly influenced by the SE lifetime. Above threshold (C, D, E), DTP induced DH is observed, truncating part of the photon pulse (see blue shaded regions), which leads to an asymmetrical pulse shape. As the pump power increases above threshold, the pump pulse crosses threshold at a smaller $t_1$ and the DH effect trims away less of the output (i.e. blue shaded regions reduce in area), leading to pulse broadening. (F) - (J) The normalized autocorrelations, $g^2(\tau)$, of the photon pulses shown in (A) - (E). The cross correlations of neighboring photon pulses are identical to the autocorrelations, but are located at $\tau_n = n/R_{rep}$. By definition, $g^2(\tau)$ pulses are symmetrical, but their FWHM is proportional to those of the output photon pulses and therefore, retain the width information of the output photon pulse. The FWHM of the pulses shown are (F) 17.32, (G) 9.86, (H) 5.38, (I) 9.94, (J) 13.38 [ns].